\pdfoutput=1
\documentclass[%
 reprint,
superscriptaddress,
bibnotes,
 amsmath,amssymb,
 aps,
prb,
]{revtex4-1}

\usepackage{graphicx}% Include figure files
\usepackage{dcolumn}% Align table columns on decimal point
\usepackage{bm}% bold math
\usepackage{color}
\begin{document}

%\preprint{APS/123-QED}

\title{A Method for Predicting Nonequilibrium Thermal Expansion Using Steepest-Entropy-Ascent Quantum Thermodynamics}% Force line breaks with \\

\author{Ryo Yamada}
\email{ryo213@vt.edu}
\affiliation{Materials Science and Engineering Department, Virginia Polytechnic Institute and State University, Blacksburg, Virginia 24061, USA}
\author{Michael R. von Spakovsky}
\email{vonspako@vt.edu}
\affiliation{Center for Energy Systems Research, Mechanical Engineering Department, Virginia Polytechnic Institute and State University, Blacksburg, Virginia 24061, USA}
\author{William T. Reynolds, Jr.}
\email{reynolds@vt.edu}
\affiliation{Materials Science and Engineering Department, Virginia Polytechnic Institute and State University, Blacksburg, Virginia 24061, USA}

\date{\today}

\begin{abstract}
Steepest-entropy-ascent quantum thermodynamics (SEAQT) is an intriguing approach that describes equilibrium and dynamic processes in a self-consistent way. To date, it has primarily been applied to gas phase systems because of the difficulty in generating the complex eigenstructures (eigenvalues and eigenfunctions) associated with solid or liquid phases. In this contribution, the SEAQT modeling is extended to the condensed phase by constructing a so-called pseudo-eigenstructure, and its applicability is demonstrated by calculating the thermal expansion of metallic silver for three cases: ($a$) stable equilibrium, ($b$) along three irreversible paths from different initial nonequilibrium states to stable equilibrium, and ($c$) along an irreversible path between two stable equilibrium states. The SEAQT framework with an anharmonic pseudo-eigenstructure predicts very reasonable values for the equilibrium thermal expansion. For the irreversible cases considered, the SEAQT approach makes it possible to predict the time-dependence of lattice relaxations from some initial state to a final state.
\end{abstract}

\pacs{Valid PACS appear here}% PACS, the Physics and Astronomy
                             % Classification Scheme.
\maketitle

\section{\label{sec:level1}INTRODUCTION}

Classical thermodynamics is arguably the foundation for investigating physical systems. However, because it is based upon uniform state variables and potentials, it does not describe kinetic behavior without invoking a governing principle beyond the first and second laws of thermodynamics. An intriguing approach to building kinetics into thermodynamic description was proposed 40 years ago \cite{hatsopoulos1976-IIb,hatsopoulos1976-III,hatsopoulos1976-IIa,hatsopoulos1976-I,beretta2005generalPhD}. The approach uses a first-principles, nonequilibrium thermodynamic-ensemble formalism to combine classical thermodynamics with quantum mechanics. Its mathematical framework, recently called steepest-entropy-ascent quantum thermodynamics or SEAQT, has seen extensive development as well as experimental validation over the last three decades (e.g., see references \cite{beretta1984quantum,beretta1985quantum,beretta2006nonlinear,beretta2009nonlinear,beretta2014steepest,montefusco2015essential,cano2015steepest,beretta2017steepest,li2016steepest,li2016generalized,li2016modeling,li2016steepest2,li2014atomistic,li2017study,li2017nonequilibrium,von2014some,li2018steepest,smith2012comparison,younis2016nonequilibrium,cano2013steepest,li2015study,li2015application}).

The SEAQT framework describes the state of a system in terms of energy and entropy, both of which are well-defined properties for any state and system\cite{gyftopoulos1997entropy,cubukcu1993thermodynamics,zanchini2014recent}. These properties are determined relative to a system's so-called eigenstructure constructed from a set of possible energy eigenlevels and a time-dependent state operator (a density or probability distribution). The energy is an expectation value of the available energies of a given state, while the entropy is viewed as a measure of load sharing among the energy.  A postulated dissipation term added to the von Neumann equation for the time evolution of a density operator introduces a nonlinear dynamics based on the principle of steepest entropy ascent and provides a link between quantum mechanics and thermodynamics. This dissipation term provides an equation of motion that captures the effects of irreversibility and makes it possible to describe nonequilibrium evolutions of system state.  

Energy eigenstructures in the SEAQT theoretical framework can be constructed from appropriate quantum mechanical models. For instance, an eigenstructure for a gas phase can be constructed by assuming gas molecules move independently with translational, rotational, and vibrational modes (the ideal gas approximation). For a condensed phase, however, the interactions between molecules cannot be neglected, because they give rise to interesting properties. These interactions complicate the energy eigenstructure for solids or liquids (and real gases as well). Furthermore, for sufficiently large systems (from the micro- to the macro-scale) and regardless of phase, infinite-dimensional state spaces are present, which make solving the equation of motion computationally intractable. To deal with the latter issue, it is possible using the density of states method developed by Li and von Spakovsky\cite{li2016steepest} to construct a so-called ``pseudo-eigenstructure'' that can be used to find practical solutions to the equation of motion and effectively explore nonequilibrium processes.

To demonstrate this approach, we apply the SEAQT framework to calculate the thermal expansion of metallic silver during an irreversible process. The presentation is organized as follows.  The basis of the SEAQT equation of motion is described in Section\;II-A, while in Sec.\;II-B, the system pseudo-eigenstructure is constructed by treating the crystal as a collection of anharmonic oscillators. Sec.\;II-C calculates the thermal expansion coefficient from the position probabilities of the oscillators determined from the pseudo-eigenstructure. Sec.\;III-A validates the approach with a comparison between the calculated (equilibrium) thermal expansion and experimental data.  Finally, Sections III-B and III-C apply the approach to investigate the lattice relaxation associated with several different non-equilibrium paths.

\section{\label{sec:level2}THEORY}

\subsection{\label{sec:level2-1}Equation of motion}
Following Beretta and others\cite{beretta2014steepest,grmela2014contact,grmela1997dynamics,montefusco2015essential,ottinger1997dynamics}, a general equation of motion for a quantum system is taken to be composed of reversible and irreversible terms. In the SEAQT framework, the form of this equation of motion for an assembly of indistinguishable particles is \cite{li2016steepest,beretta2006nonlinear,beretta2009nonlinear}
\begin{equation}
\frac{d\hat{\rho}}{dt}=-\frac{i}{\hbar}[\hat{\rho},\hat{H}]-\frac{1}{\tau(\hat{\rho})}\hat{D}(\hat{\rho}) \;, \label{equation_of_motion}
\end{equation}
where $\hat{\rho}$ is the density operator, $t$ the time, $\hbar$ the reduced Planck constant, $\hat{H}$ the Hamiltonian operator, $\tau$ the relaxation time (which is based on the speed of system state evolution in Hilbert space), and $\hat{D}$ the dissipation operator. The left-hand side of the equation and the first term on the right corresponds to the time-dependent von Neumann equation. The second term on the right is the irreversible contribution that accounts for relaxation processes in the system. This dissipation term is derived via a constrained gradient in Hilbert space \cite{li2016steepest,beretta2006nonlinear,beretta2009nonlinear} that causes the system to follow the direction of steepest entropy ascent when the energy and occupation probabilities are conserved. When $\hat{\rho}$ is diagonal in the Hamiltonian eigenvector basis, $\hat{\rho}$ and $\hat{H}$ commute and the von Neumann term in the equation of motion disappears so that Eq.\;\eqref{equation_of_motion} simplifies to \cite{li2016steepest}:
\begin{equation}
\frac{dp_j}{dt}=-\frac{1}{\tau}\frac{\begin{vmatrix} 
p_j \mathrm{ln} \frac{p_j}{n_j} & p_j & \epsilon_jp_j \\
\sum p_i \mathrm{ln} \frac{p_i}{n_i} & 1 & \sum \epsilon_i p_i \\
\sum \epsilon_i p_i \mathrm{ln} \frac{p_i}{n_i} & \sum \epsilon_i p_i & \sum \epsilon_i^2 p_i
\end{vmatrix}}{\begin{vmatrix} 
1 & \sum \epsilon_i p_i \\
\sum \epsilon_i p_i & \sum \epsilon_i^2 p_i 
\end{vmatrix}} \; ,  \label{equation_of_motion_simplified}
\end{equation}
where the $p_j$ are the diagonal terms of $\hat{\rho}$, each of which represents the occupation probability of a particle being in the $j^{th}$ energy energy level, ${\epsilon}_j$. The $n_j$ are the degeneracies of ${\epsilon}_j$. Equation\;\eqref{equation_of_motion_simplified} is a system of differential equations involving the ratio of determinants. An example application to a particularly simple system --- two-energy level particles --- can be found in reference \cite{beretta2006nonlinear}.  In general, the density operator is diagonalized when there are no quantum correlations or in classical cases \cite{spakovsky2016}. In this work, the density operator is diagonalized with respect to the energy eigenstates basis (Section\;II-B) so that Eq.\;\eqref{equation_of_motion_simplified} is applicable rather than Eq.\;\eqref{equation_of_motion}. 

When there is a heat interaction between the system and a heat reservoir at temperature $T^R$, the SEAQT equation of motion, Eq.\;\eqref{equation_of_motion_simplified}, transforms, using the concept of hypoequilibrium state, \cite{li2016steepest,li2016generalized} into
\begin{equation}
\frac{dp_j}{dt}=\frac{1}{\tau} p_j \left[ \left( -\mathrm{ln} \frac{p_j}{n_j} - \left< s \right> \right) - \beta^R \left( \epsilon_j - \left< e \right> \right) \right] \; ,     \label{equation_motion_heat}
\end{equation}
where $\beta^R=1/k_BT^R$ ($k_B$ is the Boltzmann constant), and $\left< e \right> $ and $\left< s \right>$ are the system energy and entropy, which are, respectively, defined as
\begin{equation}
e= \left< e \right>=\sum\limits_{i} \epsilon_i p_i   \label{energy}
\end{equation}
and
\begin{equation}
s= \left< s \right>=-\sum\limits_{i} p_i \mathrm{ln} \frac{p_i}{n_i} \; .  \label{entropy}
\end{equation}
where Eq.\;\eqref{entropy} is the von Neumann expression for the entropy. Provided the density operator is based on a homogeneous ensemble \cite{hatsopoulos1976-I,hatsopoulos1976-IIa,hatsopoulos1976-IIb,hatsopoulos1976-III}, this expression satisfies all the characteristics of entropy required by thermodynamics without making entropy a statistical property of the ensemble %, i.e., of a heterogeneous ensemble
\cite{gyftopoulos1997entropy,cubukcu1993thermodynamics}.

\subsection{\label{sec:level2-2} Pseudo-eigenstructure: anharmonic oscillator}
To model thermal expansion of a crystalline solid, we treat atoms in the lattice as coupled anharmonic oscillators vibrating about equilibrium positions. Unlike for the simple case of harmonic oscillators, the eigenvalues and eigenfunctions of anharmonic oscillators generally cannot be solved analytically. They can, however, be found numerically. The details of how this is done are found in reference \cite{henri2011quantum}, and the essential points are described below. 

To determine the anharmonic eigenstates, a Morse potential for the pair interaction energy is assumed, i.e., 
\begin{equation}
V_{\mbox{\scriptsize Morse}} (x) = A+D (1-e^{-\lambda (x-x_0)})^2   \label{morse_potential}
\end{equation}
where $A$, $D$, $\lambda$, and $x_0$ are fitting parameters. Since there are four fitting parameters and two of them are nonlinear, the fitting procedure is operationally difficult. The fitting procedure, however, can be simplified with the approach of Moruzzi $et$ $al$. \cite{moruzzi1988calculated} To obtain a realistic Morse potential for silver, the parameters in Eq.\;\eqref{morse_potential} are fitted to electronic total energy calculations for silver performed using the projector augmented-wave (PAW) method \cite{kresse1996efficiency} as implemented in VASP. A description of the {\it ab-initio} calculations and the resulting Morse potential parameters are provided in Appendix\;A.

Substituting the Morse potential into the time-independent Schr\"{o}dinger equation (i.e., energy eigenvalue problem),
\begin{equation}
\begin{split}
&\quad\quad\quad\quad \hat{H} \psi_n (x) = \epsilon_n \psi_n (x) \; , \\
-\frac{\hbar^2}{2m} & \frac{d^2 \psi_n (x)}{d x^2} + V_{\mbox{\scriptsize Morse}} (x)\psi_n(x) =\epsilon_n \psi_n (x) \; .  \label{schrodinger_equation}
\end{split}
\end{equation}
With the ladder operators, $a_+$ and $a_-$, the Hamiltonian can be rewritten as 
\begin{eqnarray}
\begin{split}
\hat{H}=\hbar \omega \left\{ \frac{1}{2} \left( a_-a_+ + \frac{1}{2} \right)- \frac{1}{4} \left( (a_-)^2 + (a_+)^2 \right) \right. \\ \left. + \zeta \left( 1-2\hat{F} (\lambda^0) + \hat{F} (2\lambda^0) \right) \right\} ,  \label{hamiltonian}
\end{split}
\end{eqnarray}
where
\[
\begin{array}{c c}
\hat{F}(\lambda^0)=e^{-\lambda^0 (a_- + a_+)} \; ,
&
\hat{F}(2 \lambda^0)=e^{-2 \lambda^0 (a_- + a_+)} \; , \\ \\
\zeta=\frac{D}{\hbar\omega} \; , \;\;\;\;\;\; \mbox{and}
&
\lambda^0 \; = \; \frac{1}{2}\sqrt{\frac{\hbar\omega}{D}} \; = \; \frac{1}{2}\sqrt{\frac{1}{\zeta}} \;
\end{array}
\]
The eigenfunctions, $\psi_n(x)$, for the anharmonic oscillator are then given by \cite{henri2011quantum}
 \begin{equation}
\psi_n(x)=\sum\limits_{k=0}^{n_{\mbox{\scriptsize max}}-1} C_{kn} \psi^{\mbox{\scriptsize HO}}_k(x) \; ,  \label{anharmonic_wavefunction}
\end{equation}
where the $C_{kn}$ are coefficients of the expansion, the $\psi^{\mbox{\scriptsize HO}}_n(x)$ are the eigenfunctions of a related harmonic oscillator, and $n_{\mbox{\scriptsize max}}$ is the largest quantum number of eigenvalues used for the numerical calculation. The latter is chosen to include enough energy eigenlevels to adequately represent the thermal expansion of the system with the available computational resources. The procedure for determining the harmonic oscillator eigenfunctions, $\psi^{\mbox{\scriptsize HO}}_n(x)$, and selecting $n_{\mbox{\scriptsize max}}$ is described in Appendix\;B.

The vibrational frequencies, $\omega$, in Eq.\;\eqref{hamiltonian}, are obtained from the Debye approximation\cite{kittel1966introduction} in which the velocity of sound is taken to be a constant and the lattice vibrates with frequencies up to the Debye frequency, $\omega_D$. Adopting a definition suggested by Moruzzi $et$ $al$. \cite{moruzzi1988calculated}, the constant velocity of sound at the ground state, $v_0$, is given by
\begin{equation}
v_0=0.617 \sqrt{\frac{B_0}{\rho_0}} \; ,  \label{sound_velocity}
\end{equation}
where $B_0$ and $\rho_0$ are, respectively, the bulk modulus and the density of the specimen evaluated with the lattice constant at 0\;K, $a_0$.  The coefficient, $0.617$, comes from the fact that there are two wave modes, transverse and longitudinal, whose velocities have different dependencies on the bulk modulus \cite{moruzzi1988calculated}. The bulk modulus at the ground state is given as  
\begin{equation}
\begin{split}
\quad\quad\quad B_0 & =-V_0 \left( \frac{\partial P}{\partial V} \right) _{V_0}=\frac{4}{9a_0} \left( \frac{\partial^2 E_{\mbox{\scriptsize total}}}{\partial a^2} \right ) _{a_0}  \\
& =\frac{4}{9a_0} 6 (2 D \lambda^2) \; ,  \label{bulkmodulus}
\end{split}
\end{equation}
where $P$ is the system pressure, $V$ is the volume ($V_0$ is the volume evaluated at $a_0$), $E_{\mbox{\scriptsize total}}$ is the total energy of the system, and $D$ and $\lambda$ are the fitting parameters in the Morse potential. The factor of 6 in Eq.\;\eqref{bulkmodulus} is related to the number of first-nearest-neighbor pairs (as seen in Eq.\;\eqref{pair_interaction_energy_derivation}). The Debye frequency, $\omega_{D,0}$, and the density of states, $g_0(\omega)$, evaluated at $a_0$ are given by \cite{kittel1966introduction}
\begin{equation}
\omega_{D,0}=\left( \frac{6\pi^2N_s}{V_s} \right)^{\frac{1}{3}}v_0   \label{debye_frequency}
\end{equation}
and
\begin{equation}
g_0(\omega)=\frac{9N_s\omega^2}{\omega_{D,0}^3} \; ,  \label{density_of_states}
\end{equation}
where $N_s$ and $V_s$ are, respectively, the number of primitive cells in a specimen and the volume of the specimen ($V_s=N_{s} a_0^3/4$). $N_s$ is set equal to Avogadro's number. 

The incorporation of vibrational frequencies below the Debye frequency cannot be done in a straightforward manner because the density of states is not discrete as is evident from Eq.\;\eqref{density_of_states}. To avoid this difficulty, the density of states method developed by Li and von Spakovsky \cite{li2016steepest} within the SEAQT framework is applied. This strategy combines similar energy eigenvalues into discrete bins and significantly reduces the computational burdens without losing the accuracy of the result. By using this method, which is based on a quasi-continuous assumption, the vibrational frequency and degeneracy in the pseudo-system becomes (see Appendix\;C)
\begin{equation}
\omega_i=\frac{1}{N_i}\int_{\bar{\omega}_i}^{\bar{\omega}_{i+1}}\omega g_0 (\omega)d\omega   \label{vibrational_frequency}
\end{equation}
and
\begin{equation}
N_i=\int_{\bar{\omega}_i}^{\bar{\omega}_{i+1}} g_0(\omega)d\omega \; ,  \label{degeneracy}
\end{equation}
where $\bar{\omega}_i$ is the vibrational frequency of the original system in the $i^{th}$ frequency interval. The above vibrational frequencies, Eq.\;\eqref{vibrational_frequency}, are used with Eq.\;\eqref{schrodinger_equation} to derive the eigenvalues and eigenfunctions of the crystal of anharmonic oscillators.

\subsection{\label{sec:level2-3}Thermal expansion}
Once the eigenstructure is determined, the probability of a particle at a given position and time, $\Phi(x,t)$, can be described as \cite{smith2012intrinsic}
\begin{equation}
\Phi(x,t)=\sum\limits_i p_i (t) |\psi_i (x)|^2 \; .  \label{existence_probability}
\end{equation}
The position of the oscillator, $x$, corresponds to the interatomic distance in the present application.  The expected lattice constant, $a$, at time, $t$, is then determined from the position probability of a particle, Eq.\;\eqref{existence_probability}, and its position, $x_i$, such that 
\begin{equation}
a=\langle a \rangle = \sum\limits_i \Phi(x_i, t)x_i \; .  \label{lattice_constant}
\end{equation}

The values of the occupation probabilities, $p_i$, in Eq.\;\eqref{existence_probability} reflect the distribution of the oscillators in the crystal among the various energy eigenlevels of the eigenstructure.  If a solid is at thermal equilibrium, a natural choice for the occupation probabilities is given by the canonical distribution:
\begin{equation}
P^{\mbox{ \it se}}_j=\frac{N_j e^{-\beta E_j}}{\sum\limits_i N_i e^{-\beta E_i}}=\frac{N_j e^{-\beta E_j}}{Z} \; ,  \label{canonical_distribution}
\end{equation}
where $Z$ is the partition function, $\beta$=1/$k_BT$, and $\mbox{\it se\/}$ denotes stable equilibrium. We denote the occupation probability, energy eigenvalue, and energy degeneracy are, respectively, as $P_j$, $E_j$ and $N_j$ (instead of $p_j$, $\epsilon_j$ and $n_j$) in order to emphasize that these quantities apply to the pseudo-system, which closely approximates the real system. 

\begin{figure}
\includegraphics[scale=0.58]{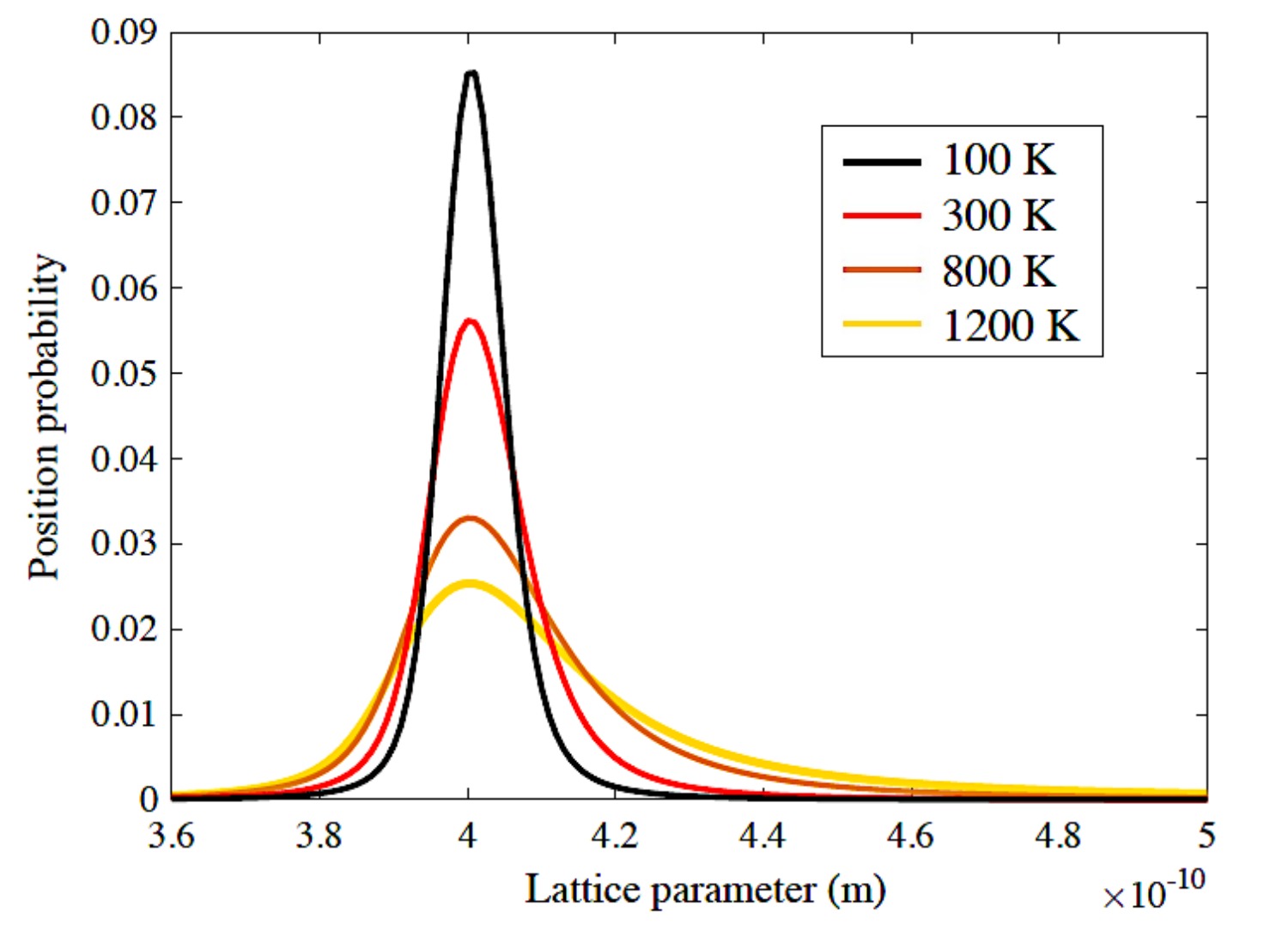}
\caption{\label{fig:existence_probability_stable_equi} The position probabilities of a particle at various temperatures calculated using the localized density approximation (LDA) functional.}
\end{figure}

Figure\;\ref{fig:existence_probability_stable_equi} shows how the occupation probability affects the position probability for silver oscillators at a series of different temperatures.  The peak in the position probabilility is the same at all temperatures and corresponds to the ground state lattice parameter (corresponding to the minimum pair interaction energy (see Fig.\;\ref{fig:pair_energy} below)). Nevertheless, the distribution becomes broader as the temperature increases since the energy is redistributed into higher lattice parameters with temperature because of the asymmetric characteristic of the potential. 

Equation\;\eqref{lattice_constant} in conjunction with the pseudo-eigenstructure and Eq.\;\eqref{canonical_distribution} provides a means to calculate the lattice constant at any temperature. The coefficient of thermal expansion, $\alpha$, is obtained from the temperature-dependent lattice constants via the relationship
\begin{equation}
\alpha(T)=\frac{a(T)-a(T_{0 \mathrm{K}})}{a(T_{0 \mathrm{K}})} \; .  \label{linear_thermal_expansion}
\end{equation}
Here, $a(T_{0 \mathrm{K}})$ includes zero-point vibrations and is different from $a_0$. The estimated lattice constant for $a(T_{0 \mathrm{K}})$, 1.002$a_0$ \cite{wu2006more}, is used here because it is difficult to satisfy the quasi-continuous condition, Eq.\;\eqref{quasi_continuous_condition},  at very low temperatures. 

It is important to note that a pseudo-eigenstructure can be constructed from any reasonable solid-state model. The one employed here for thermal expansion arises from anharmonic oscillators. It evaluates the phonon dispersion relation (or the Debye frequency) at the ground state (corresponding to $a_0$) and determines the thermal expansion from the intrinsic asymmetry of the pair-potential curve in an intuitive way. Nevertheless, a useful pseudo-eigenstructure could be built just as easily from the harmonic oscillator approximation with a volume-dependent phonon dispersion relation (i.e., the quasi-harmonic approximation; see, for example, references\cite{grabowski2007ab,moruzzi1988calculated}). 
Also, additional contributions to thermal expansion, such as electronic contributions, can be incorporated into the pseudo-eigenstructure if desired. 
% although thermal excitation of electrons is small except at exceedingly high temperatures ($\sim 10^4$ K) \cite{quong1997first,xie1999first}. 
Furthermore, the Debye model is used here for the sake of simplicity to calculate the density of states of vibrational frequencies. For more accurate calculations, a more detailed description of the density of states would be required and it can be obtained from first-principles calculations (e.g., density functional theory (DFT)).
%Although the density of states in magnetic metals is quite complex, it is much simpler in non-magnetic metals as can be seen from the phonon dispersion relations shown in reference \cite{grabowski2007ab} Thus, the assumed Debye model is a reasonable one for silver. Nevertheless, the Debye model is not a necessary assumption in the SEAQT framework. For more accurate calculations or for metals whose density of states is complex, a more detailed description of the density of states would be required and it can be obtained from first-principles calculations (e.g., DFT).

\section{\label{sec:level3}RESULTS}

\subsection{\label{sec:level3-1}Thermal expansion at stable equilibrium}
\begin{figure}
\includegraphics[scale=0.54]{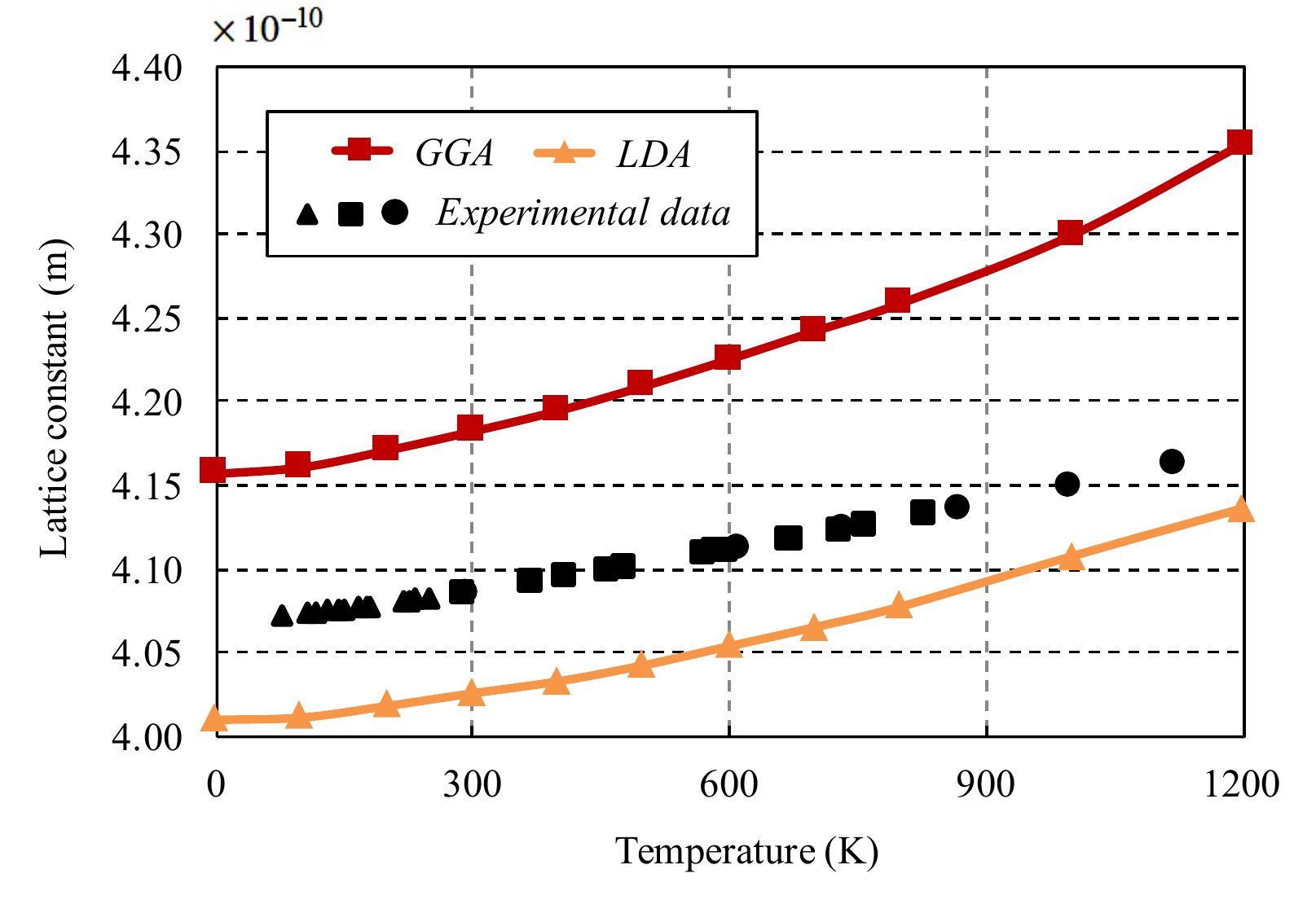}
\caption{\label{fig:temperature_dependence_lattice_constant} The temperature dependence of the Ag lattice constant. The red/yellow lines are calculated using the GGA/LDA functionals. The solid black circles, squares, and triangles are experimental data \cite{pearson1958handbook}. The lattice constant at 0\;K is estimated as $1.002a_0$, where $a_0$ is the ground state lattice constant without zero-point vibrations \cite{wu2006more}.}
\end{figure}

\begin{figure}
\includegraphics[scale=0.58]{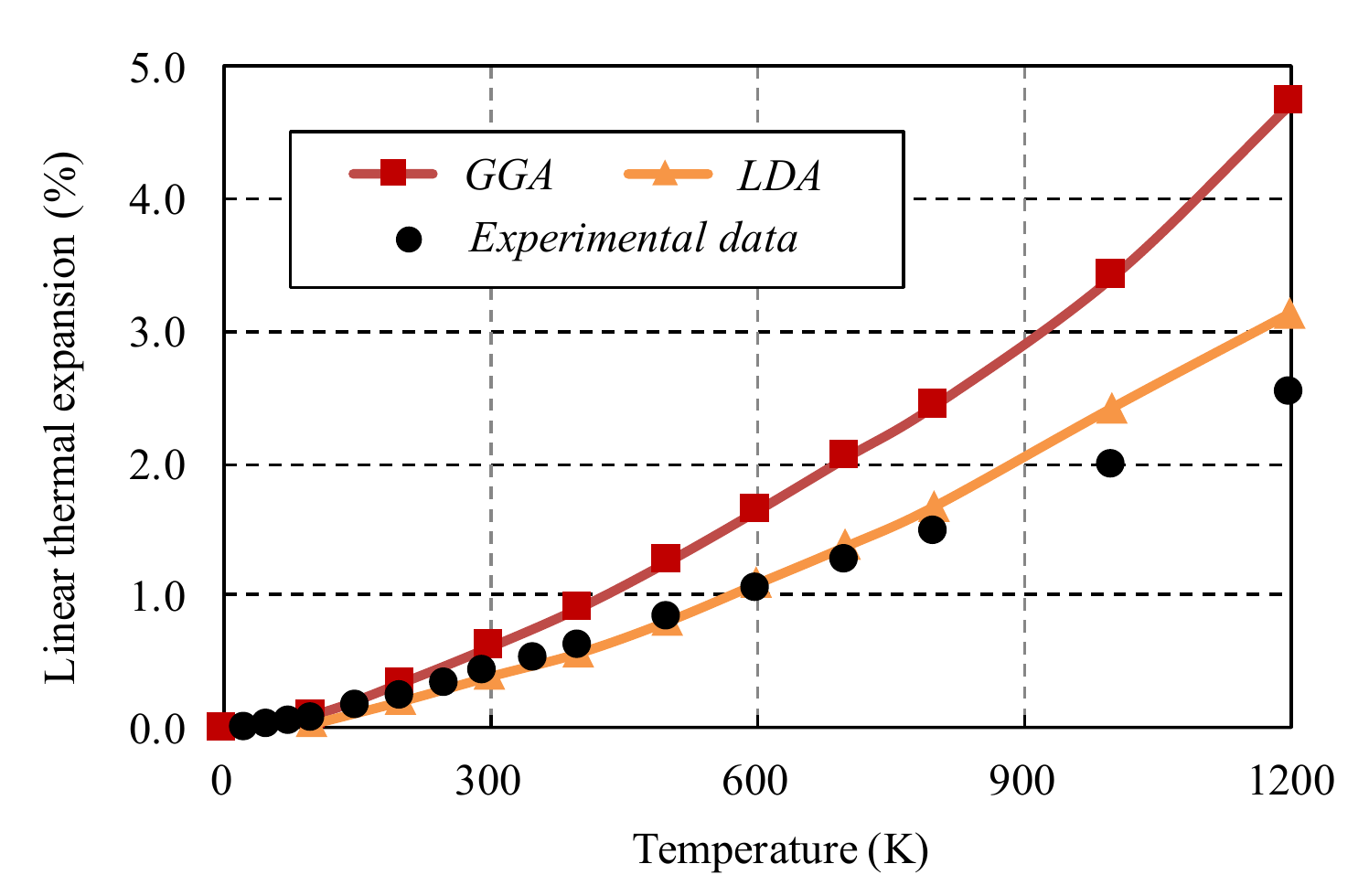}
\caption{\label{fig:linear_thermal_expansion} The temperature dependence of the linear thermal expansion coefficient. The red/yellow lines are calculated using the GGA/LDA functionals. The solid black circles are experimental data \cite{pearson1958handbook}. }
\end{figure}

Calculated equilibrium lattice constants for silver over a range of temperatures using the occupation probabilities given by Eq.\;\eqref{canonical_distribution} are shown in Fig.\;\ref{fig:temperature_dependence_lattice_constant} for two different choices of the Morse potential fitted to the total energy, i.e., one calculated with DFT using the localized density approximation (LDA) and the other using the generalized gradient approximation (GGA) approximation. The thermal expansion coefficient, Eq.\;\eqref{linear_thermal_expansion}, determined from these values is plotted in Fig.\;\ref{fig:linear_thermal_expansion}.

The solid curves in Fig.\;\ref{fig:linear_thermal_expansion} calculated from the anharmonic model with the SEAQT framework show that the thermal expansion obtained from the GGA functional is an overestimate, whereas that from LDA is quite close to the experimental results up to a fairly high temperature ($\sim$\;$800$\;K). The improved agreement of the LDA thermal expansion arises from the fact that the LDA functional overestimates both the Debye temperature and Gr\"{u}neisen constant, which are, respectively, related to the potential curvature and asymmetry/softening effect, and these errors offset each other (see Appendix\;D). Because the LDA thermal expansion is closer to experiment, only the LDA functional is used in the calculations of the following sections (Sections III-B and III-C). The qualitative agreement between experimental thermal expansion and predicted values from the SEAQT framework suggests that the pseudo-eigenstructure built from an anharmonic oscillator model is reasonable.

\subsection{\label{sec:level3-2}Nonequilibrium lattice evolution}
In this section, the SEAQT model is used to explore how the state of the silver lattice evolves from an initial, nonequilibrium state to a stable equilibrium state with an equilibrium lattice constant. An initial state can be generated in a variety of ways. Two different approaches are used here. The first selects an initial occupation distribution described by an appropriate probability distribution.  For example, Li and von Spakovsky \cite{li2016steepest, li2016generalized} generated initial states with a gamma function distribution of the form 
\begin{equation}
P^0_j = \frac{N_j E_{j}^{\theta} e^{-\beta E_j}}{\sum_i N_i E_{j}^{\theta}e^{-\beta E_i}} \; , 
\label{initial_probability_gamma}
\end{equation}
where $\beta$ is as defined above and $\theta$ represents an adjustable parameter that shifts the initial state away from the canonical distribution. In the second approach, the initial state can be generated using an {\it ad hoc} description of the initial occupation distribution.  For example, an occupation distribution generated by pumping energy into the silver lattice with a laser might be approximated by assuming that the injected photons excite silver atoms out of the lower energy states. To numerically determine such an initial state, the procedure developed by Beretta \cite{beretta2006nonlinear} is used here. The initial probability distribution, $P^0_j$, is expressed in terms of the following perturbation function:
\begin{equation}
P^0_j=(1-\lambda_{\mbox{\scriptsize const}})P^{\mbox{\footnotesize \it pe}}_j+\lambda_{\mbox{\scriptsize const}}P^{\mbox{\footnotesize  \it se}}_j \; ,  \label{initial_probability_partial}
\end{equation}
where $P^{\; \mbox{\footnotesize \it pe/se}}_j$ are the partially canonical and stable equilibrium probability distributions and $\lambda_{\mbox{\scriptsize const}}$ is the perturbation constant (assumed to be 0.1 here) that describes the initial departure from the partially canonical state. For the partially canonical distribution, it is assumed that the atoms do not occupy the lowest three quantum levels. 

Now, in order to determine the maximum vibrational quantum number and apply the quasi-continuous condition of the density of states method \cite{li2016steepest}, Eqs.\;\eqref{convergence_condition_phonon} and \eqref{quasi_continuous_condition} given in Appendices B and C must be satisfied at stable equilibrium as well as at nonequilibrium. Both conditions are satisfied rigorously at stable equilibrium but require an additional concept at nonequilibrium, namely, that of hypoequilibrium state \cite{li2016steepest}. With this concept, Eqs.\;\eqref{convergence_condition_phonon} and \eqref{quasi_continuous_condition} can be satisfied rigorously at nonequilibrium. A discussion of this concept is beyond the scope of the present paper and, thus, the reader is referred to reference \cite{li2016steepest}. 
%Furthermore, if the initial nonequilibrium state is at a lower energy than that of the final stable equilibrium state, Eq.\;\eqref{convergence_condition_phonon} is satisfied, while Eq.\;\eqref{quasi_continuous_condition} is satisfied when the nonequilibrium state is generated from a gamma distribution.

\begin{figure}
\includegraphics[scale=0.56]{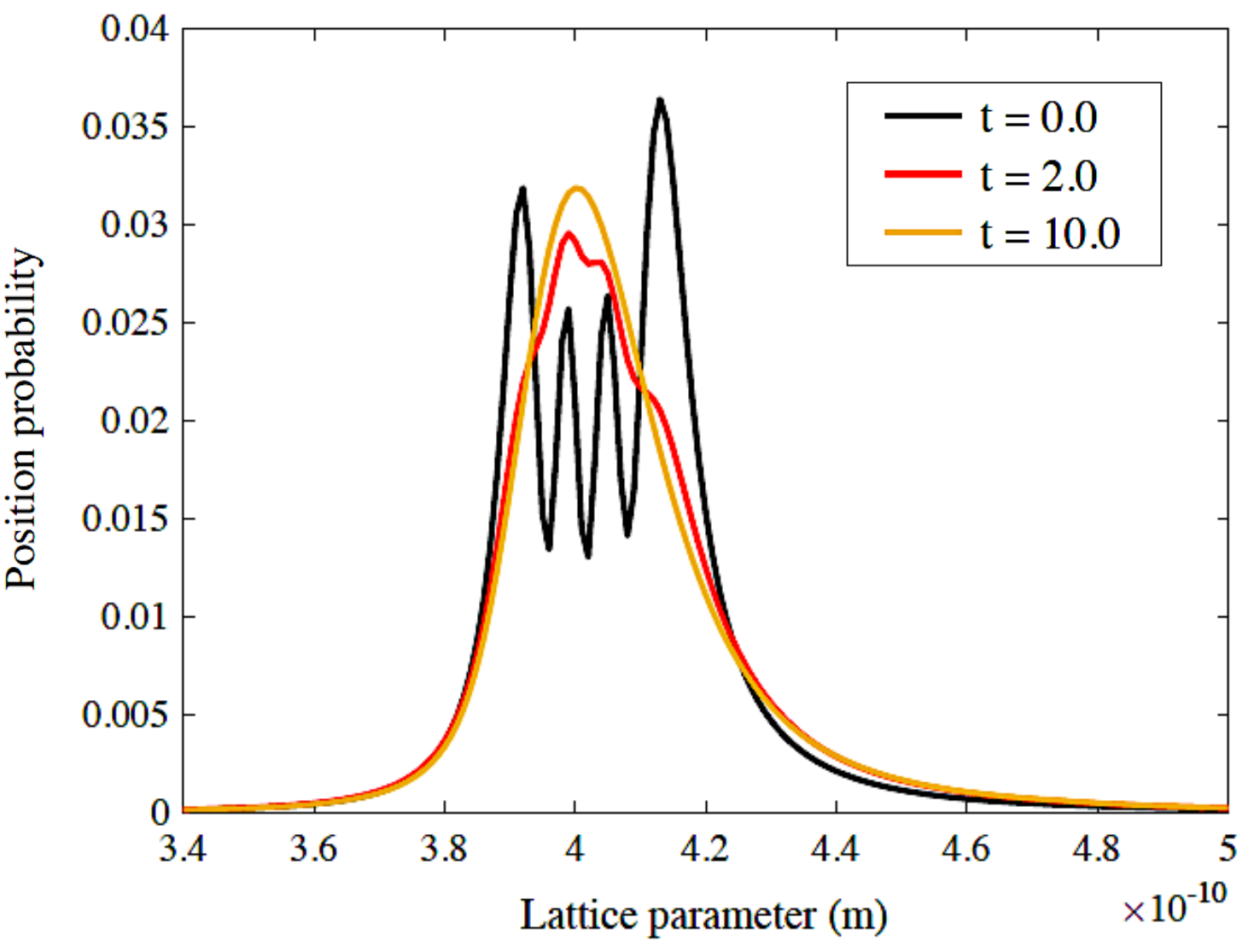}
\caption{\label{fig:probability_nonequi_isolated} The position probability distribution of a particle in an isolated system as a function of lattice parameter at three different times of the irreversible thermodynamic path determined using the initial state generated from Eq.\;\eqref{initial_probability_partial} and the LDA functional. The temperature in the final stable equilibrium state is 800\;K.}
\end{figure}

\begin{figure}
\includegraphics[scale=0.49]{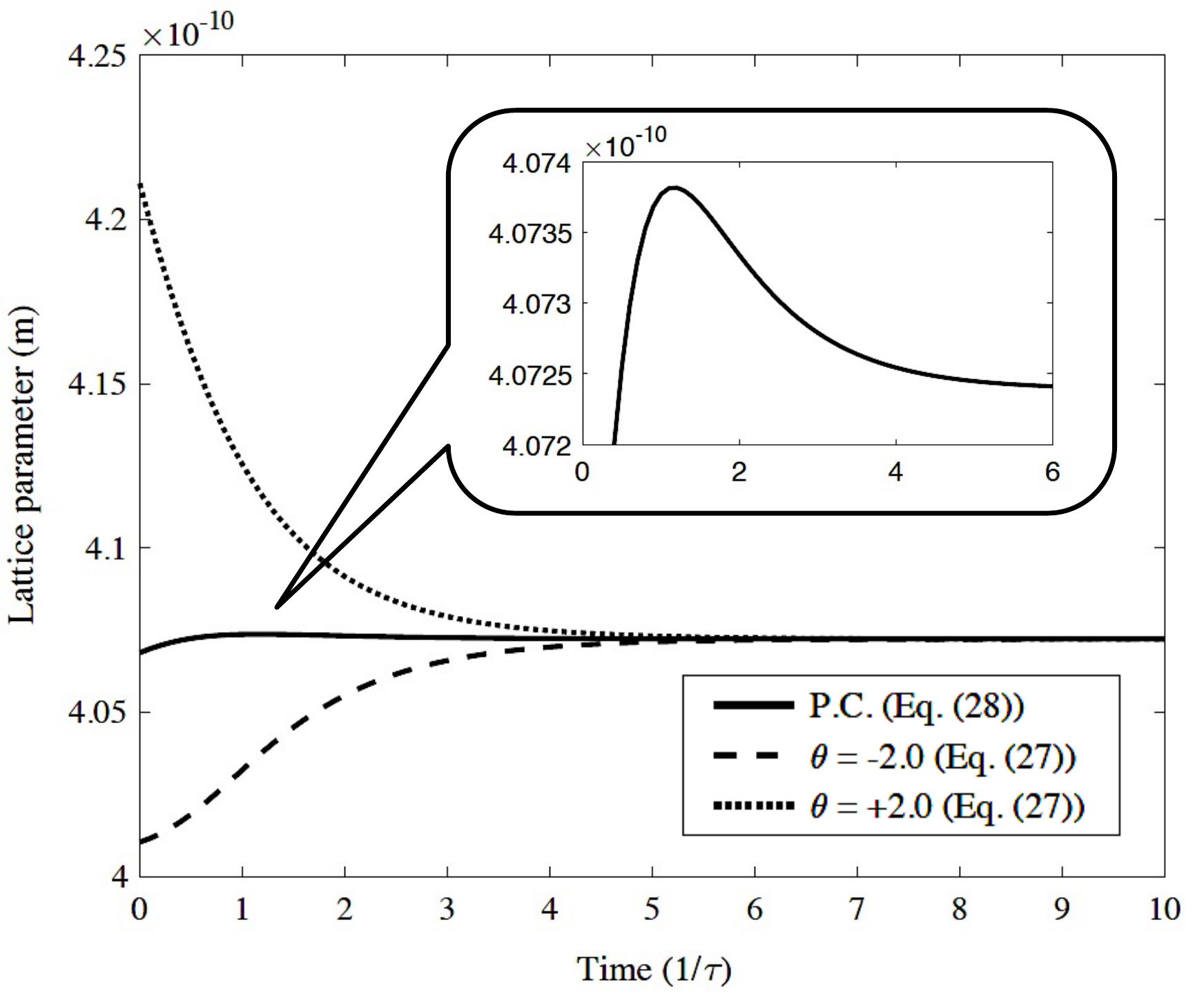}
\caption{\label{fig:time_lattice_parameter_isolated} The time dependence of the Ag lattice parameter for three initial states using the LDA functional. In each case, the lattice parameter relaxes to the stable equilibrium value at 800\;K (Figure\;\ref{fig:temperature_dependence_lattice_constant}). The insert shows the non-monotonic behavior of the lattice parameter for the evolution based on the initial state prepared by the partially canonical (P.C.) distribution, Eq.\;\eqref{initial_probability_partial}.  }
\end{figure}

Fig.\;\ref{fig:probability_nonequi_isolated} shows an example of how the position probability distribution varies with lattice parameter at three different times for the irreversible thermodynamic path determined for the initial state generated using the partially canonical distribution of Eq.\;\eqref{initial_probability_partial}. Figure\;\ref{fig:time_lattice_parameter_isolated} shows the time dependence of the silver lattice parameter for three different initial states, one generated based on Eq.\;\eqref{initial_probability_partial} and two based on Eq.\;\eqref{initial_probability_gamma}. For the former, the change in the lattice parameter is quite small because the total energy remains fixed throughout the evolution. For the other two, the system energy varies since the system is allowed to interact with a heat reservoir at 800\;K. All three evolutions arrive at the same stable equilibrium state of 800\;K. An interesting feature of the evolution based on Eq.\;\eqref{initial_probability_partial} is that a non-monotonic change of the lattice parameter with time is observed. Note that the lattice parameter at stable equilibrium for all three evolutions corresponds to the lattice constant at 800\;K derived from the canonical distribution in Sec.\;III-A (see Fig.\;\ref{fig:temperature_dependence_lattice_constant}).

In these results, the time scales are normalized by the relaxation time, $\tau$, which can be correlated to the real time, $t$, via comparisons to experimental data \cite{beretta2017steepest,li2015study,li2017nonequilibrium} or from $ab$ $initio$ calculations based on quantum or classical mechanical considerations \cite{beretta2014steepest,li2016generalized,li2018steepest}. The latter approach is used in the following section (Sec.\;III-C). Nonetheless, the unique kinetic path predicted by the SEAQT framework is independent of the value of the relaxation time assumed, since the path in state space remains the same. The relaxation time $\tau$ simply determines how fast along that path the state of the system changes.

\subsection{\label{sec:level3-3}Thermal expansion along irreversible path between equilibrium states}
\begin{figure}
\includegraphics[scale=0.45]{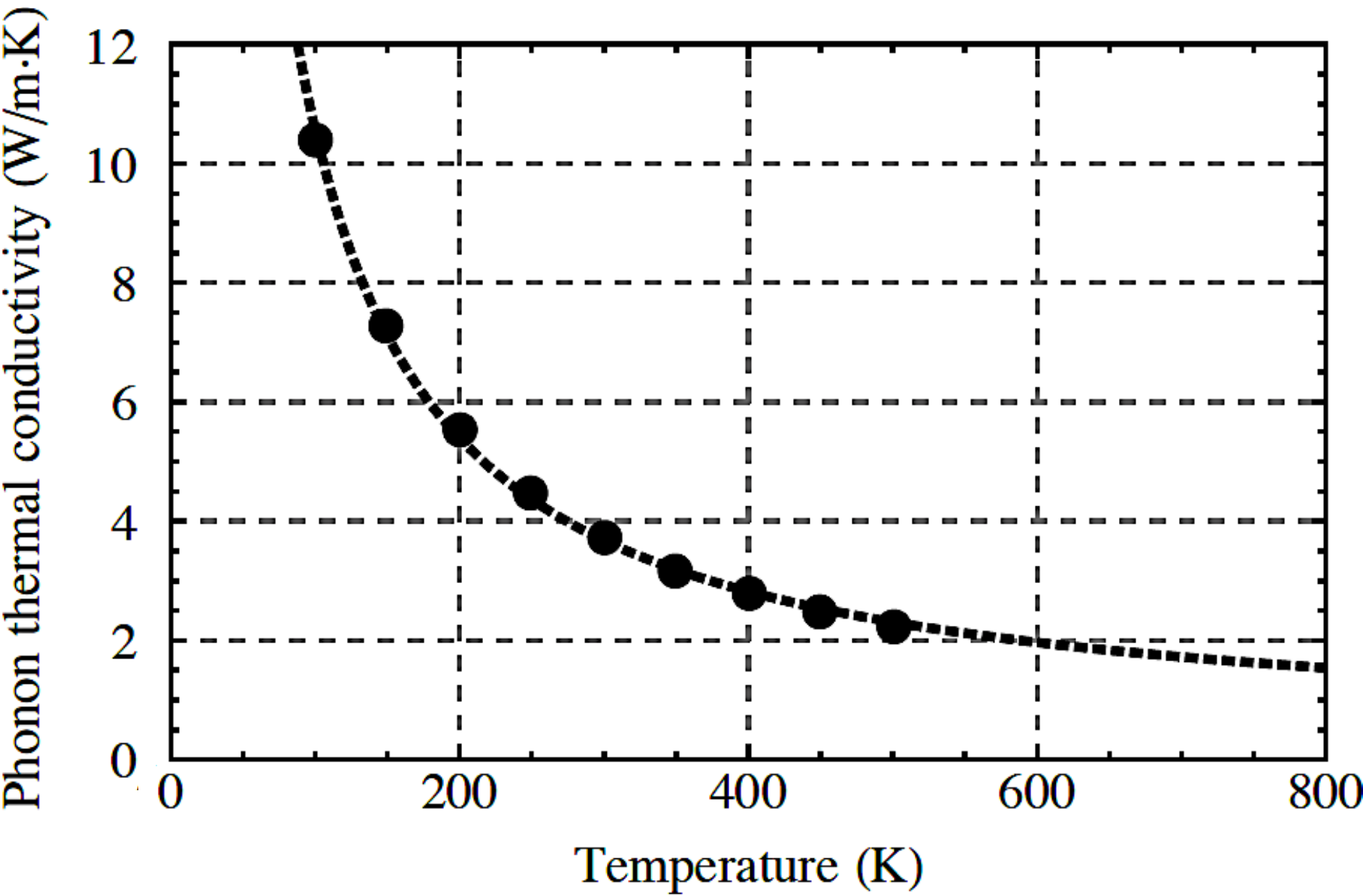}
\caption{\label{fig:thermal_conductivity} The temperature dependence of the phonon thermal conductivity coefficient. The black circles are the original data  \cite{jain2016thermal}, and the broken line is the fitting function, $\ K_{\mbox{\scriptsize phonon}}(T) = A'+B' \exp(C'/T)$, where $A'=-504.3$, $B'=504.6$, and $C'=2.020$.}
\end{figure}

The real-time dependence of the lattice parameter can be estimated by correlating the relaxation time, $\tau$, to the phonon contribution of thermal conductivity. This is done by allowing the system to interact with a heat reservoir in an irreversible process in which the system starts at an initial temperature $T_0$ and ends up in thermal equilibrium with the reservoir at $T^R$. The calculated time dependence of the lattice parameter for $T_0=300$\;K and $T^R=800$\;K is shown in Fig.\;\ref{fig:time_lattice_parameter_heat}. Here, the real time scale is determined using the following equation (see Appendix\;F):
\begin{equation}
\ \tau = \frac{N}{4LK_{\mbox{\scriptsize phonon}}} \frac{1}{T^R-T^C}\frac{dE}{dt^*} \; , \label{relaxation_time}
\end{equation}
where $N$ is the number of atoms in a sample, $L=10$\;mm is the length of the assumed cubic sample, $K_{\mbox{\scriptsize phonon}}$ is the phonon component of the thermal conductivity coefficient, $T^C$ is the temperature at the center of the sample, and $dE/dt^*$ is the energy change rate per atom, which can be determined from the SEAQT framework. The time dependence of $T^C$ can be estimated using the relation between the lattice constant and temperature derived in the stable equilibrium calculation of Sec.\;III-A (Fig.\;\ref{fig:temperature_dependence_lattice_constant}). The temperature dependence of the phonon thermal conductivity coefficient for silver calculated taking into account the electron-phonon and phonon-phonon interactions from first-principles by Jain $et$ $al$. \cite{jain2016thermal} is employed using the following equation: 
\begin{equation}
K_{\mbox{\scriptsize phonon}}(T) = A'+B' \exp(\frac{C'}{T}) \; , \label{phonon_fitting}
\end{equation}
where $T$ is temperature, and $A'$, $B'$, and $C'$ are fitting parameters. The fitting result is shown in Fig.\;\ref{fig:thermal_conductivity} with the original data \cite{jain2016thermal}. Eq.\;\eqref{phonon_fitting} closely reproduces the original values although the reliability of the fitted function above 500\;K can not be verified because there is no calculated data above this temperature. 
\begin{figure}
\includegraphics[scale=0.5]{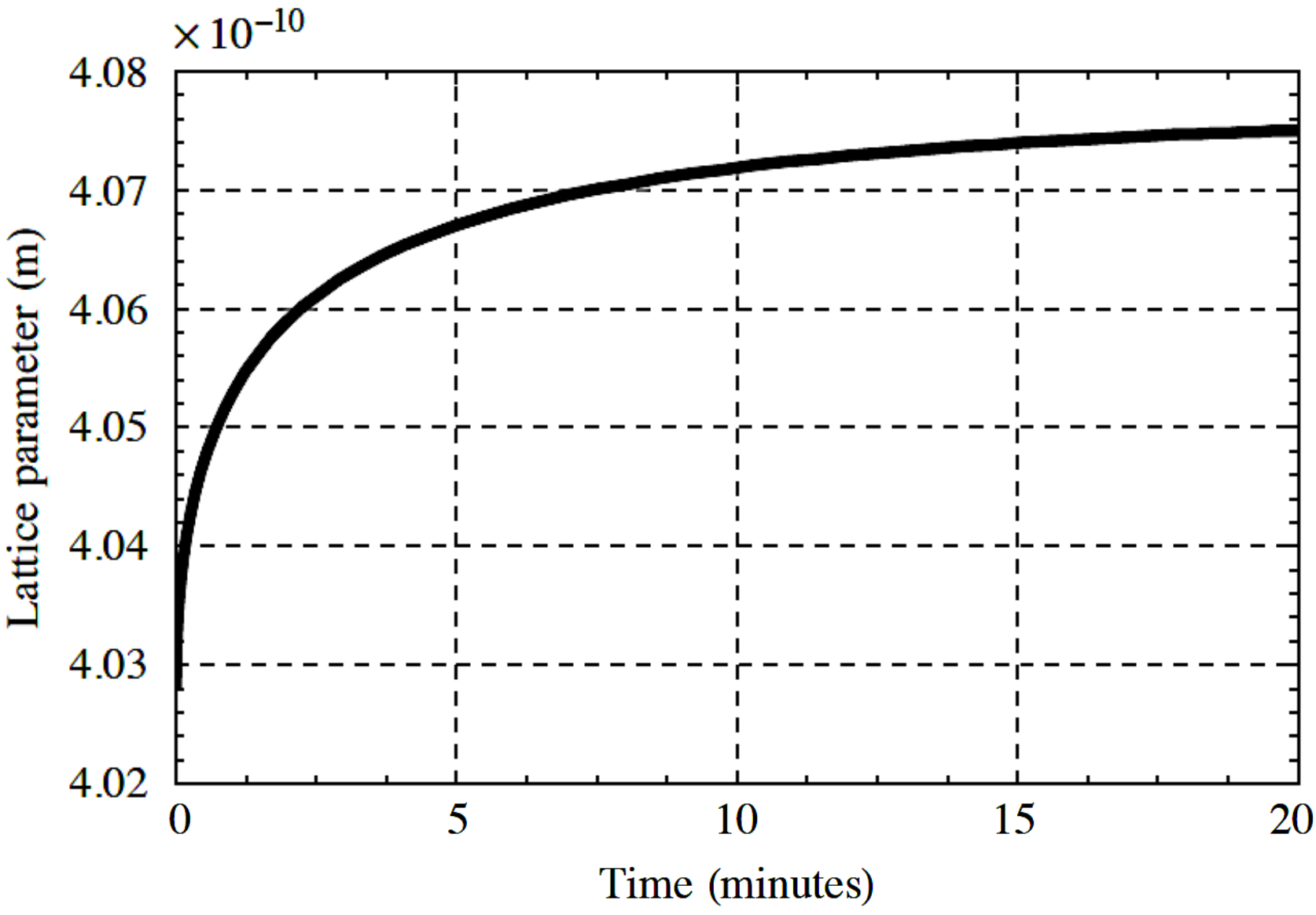}
\caption{\label{fig:time_lattice_parameter_heat} The time dependence of the lattice parameter in a cubic sample ($L=10$\;mm) interacting with a heat reservoir ($T^R=800$\;K) using the SEAQT framework and the LDA functional. }
\end{figure}

From Fig.\;\ref{fig:time_lattice_parameter_heat}, it is apparent that the steepest part of the lattice change happens at the first half of the relaxation process (i.e., from 0\;$\sim$\;10 minutes) and slows as the temperature of the sample, $T^C$, approaches the temperature of the reservoir, $T^R$.

\section{\label{sec:level1} CONCLUSIONS}
The SEAQT model can reliably estimate the thermal expansion of face-centered-cubic silver with a pseudo-eigenstructure based on a crystal of anharmonic oscillators. The accuracy of the approach is quite good for anharmonic oscillators determined from band calculations. It has the added advantage that the thermal expansion can be calculated not only at stable equilibrium but also along a path from some initial non-equilibrium state to stable equilibrium.

In the present work, the following three calculations are provided: ($a$) at stable equilibrium, ($b$) along three irreversible paths from different initial non-equilibrium states to stable equilibrium, and ($c$) along an irreversible path between two stable equilibrium states. For each calculation, it is confirmed that

($a$) the SEAQT framework with an anharmonic pseudo-eigenstructure predicts reasonable values for equilibrium thermal expansion; %the linear thermal expansion of silver calculated with SEAQT and an anharmonic pseuodo-eigenstructure is in good agreement with experimental data; %and past calculations based on the quasi-harmonic approximation at low temperature;

($b$) the time dependence of the lattice parameter has a non-monotonic behavior for one particular choice of initial state prepared by a method that uses partial occupation probabilities (Eq.\;\eqref{initial_probability_partial}) to approximate energy injection from a laser. A lattice parameter that monotonically increases or decreases with time can result from initial states prepared using Eq.\;\eqref{initial_probability_gamma};

($c$) the real-time dependence of the lattice parameter is found using the phonon component of thermal conductivity and shows that the most significant lattice change occurs at the earlier stages of the relaxation process.

It is noteworthy that the SEAQT model with an anharmonic pseudo-eigenstructure based on a coupled anharmonic oscillator is much more computationally efficient than quasi-harmonic models. The latter require volume dependent phonon dispersion relations which are computationally demanding using DFT, but our approach evaluates the phonon dispersion relation only at the ground state and thermal expansion is calculated from the pseudo-eigenstructure by solving the time-independent Schr\"{o}dinger equation of anharmonic oscillators. This involves solving a modest size system of ordinary differential equations --- a comparatively small computational problem.

\section*{ACKNOWLEDGEMENT}
We acknowledge the National Science Foundation (NSF) for support through Grant DMR-1506936 and Advanced Research Computing at Virginia Tech for providing computational resources and technical support that have contributed to the results reported within this paper. URL: http://www.arc.vt.edu .

\section*{APPENDIX A: Ab-initio calculations}
To obtain a realistic Morse potential for silver, the parameters in Eq.\;\eqref{morse_potential} are fitted to electronic total energy calculations for silver performed using the projector augmented-wave (PAW) method \cite{kresse1996efficiency} as implemented in VASP. For the exchange-correlation functional, both the localized density approximation (LDA) of Ceperley and Alder \cite{ceperley1980ground,perdew1981self} and the generalized gradient approximation (GGA) of Perdew-Burke-Ernzerhof (PBE) \cite{perdew1996generalized} are employed. Supercells in the present DFT calculations contain 4 atoms in the face-centered cubic (fcc) structure, and the plane wave cut-off energy is set to 400\;eV. Integration over the Brillouin zone is done with $11$$\times$$11$$\times$$11$ $k$-points, and the tetrahedron method \cite{blochl1994improved} is applied for the $k$-space integrals. 

\begin{figure}
\includegraphics[scale=0.52]{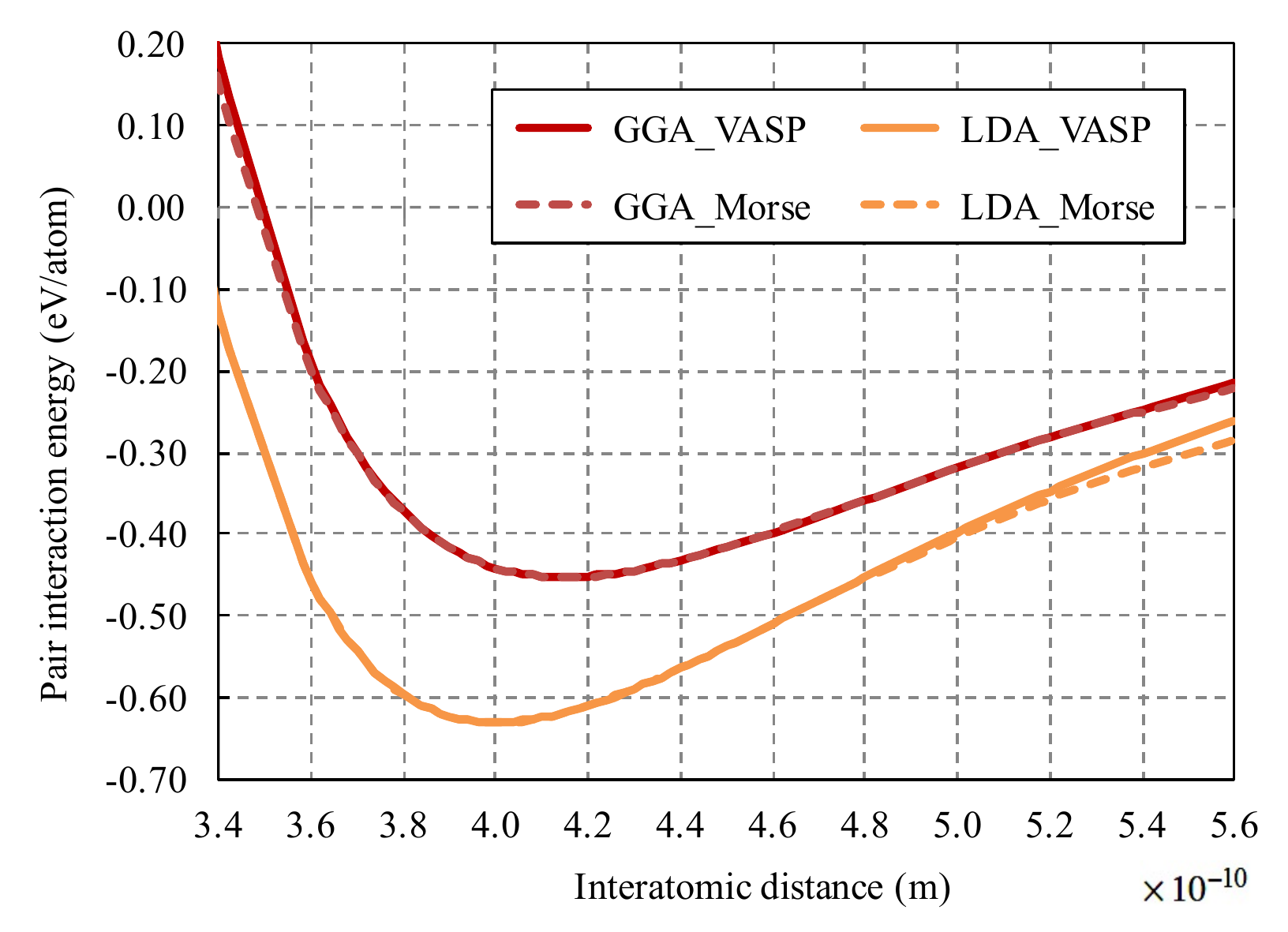}
\caption{\label{fig:pair_energy} Pair interaction energies for the GGA and LDA functionals. The solid lines are the results derived from band calculations, and the broken lines are the Morse potential function fitted to the band calculations.}
\end{figure}

The pair interaction energy, $E_{\mbox{\scriptsize pair}}$, employed in the Morse potential, Eq.\;\eqref{morse_potential}, is extracted from the total energies, $E_{\mbox{\scriptsize total}}$, derived from the band calculation by considering only first-nearest-neighbors, i.e.,
\begin{equation}
\begin{split}
E_{\mbox{\scriptsize total}}= \frac{N_c z}{2}E_{\mbox{\scriptsize pair}} \Rightarrow E_{\mbox{\scriptsize pair}}=\frac{1}{6N_c}E_{\mbox{\scriptsize total}} \; ,  \label{pair_interaction_energy_derivation}
\end{split}
\tag{A1}
\end{equation}
where $N_c$ is the number of atoms in the supercell and $z$ is the coordination number ($z=12$ for the fcc structure). The pair potentials are fitted in the range from 3.8 to 5.2\;$\mathrm{\AA}$ for GGA and 3.6 to 4.8\;$\mathrm{\AA}$ for LDA. The pair interaction energy and fitting parameters are, respectively, shown in Fig.\;\ref{fig:pair_energy} and Table\;\ref{table:morse_fitting} for the GGA and the LDA functionals. 
\begin{table}
\begin{center}
\caption{\label{table:morse_fitting}Morse potential parameters used to fit band calculations made with the GGA and LDA functionals. Note that the fitting parameter, $x_0$, corresponds to the lattice constant, $a_0$. \\}
{\renewcommand\arraystretch{1.5}
\begin{tabular}{ c  c  c } \hline \hline
$\quad\quad\quad$  & \quad \quad GGA \quad \quad & \quad \quad LDA \quad \quad \\ \hline
\quad $A$\quad($\mathrm{eV/atom}$) & -0.0914 & -0.1476 \\
\quad $D$\quad($\mathrm{eV/atom}$) & 0.3616 & 0.4829 \\
\quad $\lambda$\quad($1/\mathrm{\AA}$) & 1.1069 & 1.1648 \\
\quad $x_0$\quad($\mathrm{\AA}$) & 4.1493 & 4.0025 \\ \hline \hline
\end{tabular}
}
\end{center}
\end{table}

\section*{APPENDIX B: The  Maximum vibrational quantum number and the Harmonic Oscillator}

The vibrational quantum number can be infinite. Thus, the maximum value, which does not underestimate the thermal expansion of the lattice, must be determined. A value as small as possible that does not change the result is sought. The probability of phonons, $X_n$, whose quantum number is labeled as $n$, can be derived from the Maxwell-Boltzmann distribution as \cite{kittel1966introduction}
\begin{equation}
X_n=\frac{N_n}{\sum\limits_{s=0}^{\infty}N_s}=\frac{\mathrm{exp} (-\frac{n\hbar\omega}{k_BT})}{\sum\limits_{s=0}^{\infty} \mathrm{exp} (-\frac{s\hbar\omega}{k_BT})} \; ,  \tag{B1}\label{probability_of_phonon}
\end{equation}
where $N_n$ is the number of phonons with quantum number $n$. Checking whether or not the summation of the above probabilities from $n=0$ to $n_{\mbox{\scriptsize max}}$ is approximately one determines the maximum vibrational value, $n_{\mbox{\scriptsize max}}$, i.e., when 
\begin{equation}
\delta=1-\sum\limits_{i=0}^{n_{\mbox{\scriptsize max}}}X_i\approx0 \; .   \tag{B2}\label{convergence_condition_phonon}
\end{equation}
In this work, $n_{\mbox{\scriptsize max}}$ is set in order to satisfy the condition $\delta$\;$\textless$\;0.01. 

To determine the eigenvalues, ${\epsilon}_n^{\mbox{\scriptsize HO}}$, and eigenfunctions at a position $x$, ${\psi}^{\mbox{\scriptsize HO}}_n(x)$, for the quantum {\it harmonic} oscillator, the following analytical expressions are used \cite{griffiths2005introduction}:
\begin{equation}
\epsilon_n^{\mbox{\scriptsize HO}}=\left( n+\frac{1}{2} \right ) \hbar \omega  \tag{B3}\label{HO_eigenvalue}
\end{equation}
and
\begin{equation}
\psi^{\mbox{\scriptsize HO}}_n (x)=\left( \frac{m\omega}{\pi\hbar} \right )^{\frac{1}{4}} \frac{1}{\sqrt{2^n n!}}H_n (\xi) e^{-\frac{\xi^2}{2}} \; ,  \tag{B4}\label{HO_wavefunction}
\end{equation}
where $n$ is the quantum number ($n=0, 1, 2, \ldots$), $m$ the mass of a particle, $\omega$ the vibrational frequency, $H_n$ a Hermite polynomial, and $\xi=\sqrt{\frac{m\omega}{\hbar}}x$.

\section*{APPENDIX C: Quasi-continuous condition}
The quasi-continuous condition can be derived by seeking a condition where the properties between the original and pseudo systems become approximately the same \cite{li2016steepest}. However, past work has applied this method to the density of states for energy eigenstructures and not to that for the vibrational frequency. In this appendix, the quasi-continuous condition is extended to the density of states for this frequency. 

The quasi-continuous condition for the density of states of the energy is given as
\begin{equation}
\frac{1}{\beta}\gg |E_{i+1}-E_i| \; ,  \tag{C1}\label{quasi_continuous_condition}
\end{equation}
where $E_i$ is $i^{th}$ eigenvalue in the pseudo-system expressed as
\begin{equation}
E_i=\frac{1}{N_i}\sum\limits_j n_j^i \epsilon_j^i \; ,   \tag{C2}\label{energy_pseudo_system_discrete}
\end{equation}
where $\epsilon^i_j$ is the $j^{th}$ eigenvalue in the original system in the $i^{th}$ energy interval and $n^i_j$ is its degeneracy. The degeneracy of $E_i$, $N_i$, is defined as 
\begin{equation}
N_i=\sum\limits_j n_j^i \; .   \tag{C3}\label{degeneracy_pseudo_system_discrete}
\end{equation}
The energy intervals can then be determined as
\begin{equation}
\Delta E=\frac{1}{R} (\epsilon_{\mbox{\scriptsize cut}}-\epsilon_{\mbox{\scriptsize ground}}) \; ,   \tag{C4}\label{energy_intervals}
\end{equation}
where $\epsilon_{\mbox{\scriptsize cut}}$ and $\epsilon_{\mbox{\scriptsize ground}}$ are, respectively, the cutoff and ground state energies and $R$ is the number of intervals. For simplicity, an harmonic oscillator is assumed here to derive the analogous expression for the quasi-continuous condition of the vibrational frequency. For the harmonic oscillator, the vibrational frequency is directly related to the eigenenergy as shown in Eq.\;\eqref{HO_eigenvalue}. Thus, Eqs.\;\eqref{quasi_continuous_condition} to \eqref{energy_intervals} can be rewritten as
\begin{equation}
\frac{1}{\beta} \gg \hbar \bigg| \left( \langle n \rangle _{\omega_{i+1}} + \frac{1}{2} \right) \omega_{i+1} - \left( \langle n \rangle _{\omega_{i}} + \frac{1}{2} \right) \omega_{i} \bigg| \; , \tag{C5}\label{quasi-continuous_condition_phonon}
\end{equation}
\begin{equation}
\omega_i = \frac{1}{N_i} \int_{\bar{\omega}_i}^{\bar{\omega}_{i+1}}\omega g(\omega) d\omega \; ,  \tag{C6}\label{vibrational_frequency_pseudo_system}
\end{equation}
\begin{equation}
N_i=\int_{\bar{\omega}_i}^{\bar{\omega}_{i+1}} g(\omega) d\omega \; ,  \tag{C7}\label{degeneracy_pseudo_system_continuous}
\end{equation}
and
\begin{equation}
\Delta \bar{\omega}=\frac{\omega_D}{R} \; .  \tag{C8}\label{vibraitonal_frequency_interval}
\end{equation}
where $\omega_i$ and $N_i$ are vibrational frequency and its degeneracy in the pseudo-system, $\langle{n}\rangle_{\omega_i}$ is the average number of phonons with vibrational frequency $\omega_i$, $\bar{\omega}$$_i$ is the vibrational frequency in the original system, and $\omega_D$ is the Debye frequency. Since the Debye frequency is the maximum frequency with which the lattice can vibrate, it can be regarded as the cutoff energy (cutoff vibrational frequency). The average number of phonons at temperature, $T$, can be expressed as \cite{kittel1966introduction}
\begin{equation}
\langle n \rangle_\omega = \frac{1}{\mathrm{exp}(\hbar\omega / k_BT)-1} \; .  \tag{C9}\label{average_number_of_phonon}
\end{equation}
The above quasi-continuous condition is valid at stable equilibrium but can be extended to nonequilibrium via the hypoequilibrium concept \cite{li2016steepest}.

\section*{APPENDIX D: Analysis of GGA/LDA thermal expansion}
The origin of the difference between the calculated thermal expansion and that given by experiments (Fig.\;\ref{fig:linear_thermal_expansion}) can be inferred from three characteristics of the interatomic potential energy: its curvature, its asymmetry about the minimum bonding energy, and its width at large displacements from the minimum point (i.e., the softening effect \cite{kittel1966introduction}). Note that the first and third contributions do not have an impact on the thermal expansion for the symmetry potential of the harmonic oscillators, but do strongly affect it for the asymmetry potential. The Debye temperature, $\Theta_D$, and the Gr\"{u}neisen constant, $\gamma$, can be used to evaluate the magnitude of each of these contributions. These constants evaluated at $a_0$ are obtained from band calculations using the following relationships: \cite{moruzzi1988calculated,kittel1966introduction}
\begin{equation}
\Theta_{D,0}=\frac{\hbar \omega_{D,0}}{k_B}    \tag{D1}\label{debye_temperature}
\end{equation}
and
\begin{equation}
\gamma_0 \equiv -\left( \frac{\partial \mathrm{ln} \Theta_D}{\partial \mathrm{ln}V} \right)_{V_0} =\frac{\lambda a_0}{2} \; ,  \tag{D2}\label{Gruneisen_constant}
\end{equation}
where $\omega_{D,0}$ is given in Eq.\;\eqref{debye_frequency}. The calculated values of these constants are shown in Table\;\ref{table:Debye_Gruneisen} and compared with experimental results \cite{gschneidner1964solid,smith1995low}. The normalized Gr\"{u}neisen constant, $\gamma_0'=\gamma_0/a_0$, is used because $\gamma_0$ depends on $a_0$ ($x_0$), which does not contribute to the thermal expansion (see Appendix\;E; hereafter, we call the normalized constant, simply the Gr\"{u}neisen constant). In general, the lattice expansion becomes smaller or larger when it has a larger or smaller Debye temperature and a smaller or larger Gr\"{u}neisen constant (Appendix\;E). As seen in Table\;\ref{table:Debye_Gruneisen}, the Debye temperature and Gr\"{u}neisen constant are overestimated by the LDA functional, while only the Gr\"{u}neisen constant is overestimated by the GGA functional. For this reason, it is concluded that the GGA functional overestimates the thermal expansion because it does not accurately represent the asymmetry/softening effect in the potential energy. The closer agreement of the LDA thermal expansion to experimental values arises from the fact that the LDA functional overestimates both the Debye temperature and the Gr\"{u}neisen constant and these errors offset each other. 
\begin{table}
\begin{center}
\caption{\label{table:Debye_Gruneisen}The Debye temperature, $\Theta_{D,0}$, and Gr\"{u}neisen constant, $\gamma_0$, for Ag evaluated at $a_0$ from the GGA and LDA functionals with experimental data. \cite{gschneidner1964solid,smith1995low} The normalized Gr\"{u}neisen constant, $\gamma_0'=\gamma_0/a_0$, is shown as well. The lattice constant in Table\;\ref{table:morse_fitting} ($a_0^{\mathrm{GGA}}=4.149$\;$\mathrm{\AA}$ and $a_0^{\mathrm{LDA}}=4.003$\;$\mathrm{\AA}$) and Ref.\cite{pearson1958handbook} ($a_0^{\mathrm{Exp}}=4.076$\;$\mathrm{\AA}$) are used, respectively, for the normalization.\\ }
{\renewcommand\arraystretch{1.5}
\begin{tabular}{ c  c  c  c } \hline \hline
 & \quad GGA  \quad & \quad LDA  \quad & \quad \quad Exp. \quad \quad \\ \hline
\quad $\Theta_{D,0}$(K)  \quad & 221.9 & 266.9 &226.5 \cite{smith1995low} \\
\quad Error in  $\Theta_{D,0}$ \quad & \quad -2.03\% \quad & \quad +17.8\% \quad &  - \\ \hline
\quad $\gamma_0$ \quad & 2.623 & 2.664 & 2.46 \cite{gschneidner1964solid} \\
\quad $\gamma_0'$ $\mathrm{(1/\AA)}$ \quad & 0.6321 & 0.6656 & 0.603 \\ 
\quad Error in $\gamma_0'$ \quad & +4.74\% & +10.4\% & - \\ \hline \hline
\end{tabular}
}
\end{center}
\end{table}

\section*{APPENDIX E: The relationship of the Debye temperature and Gr\"{u}neisen constant to thermal expansion}
A simple anharmonic potential is given as \cite{henri2011quantum,kittel1966introduction}
\begin{equation}
V(x)=V_0+c(x-x_0)^2-g(x-x_0)^3+f(x-x_0)^4 \; , \tag{E1}\label{simple_anharmonic_potential}
\end{equation}
where $V_0$ is the energy at the equilibrium point, $x_0$, and $c$, $g$, and $f$ are all positive coefficients. The second and fourth terms in this potential, respectively, are related to the inverse of the curvature and the width of the potential at large amplitudes, while the third term represents the asymmetry \cite{kittel1966introduction}. It is expected that a thermal expansion becomes smaller or larger when $c$ and $f$ in the potential are smaller or larger and $g$ in the potential is larger or smaller. Furthermore, a Taylor series expansion of the Morse potential (Eq.\;\eqref{morse_potential}) is written as
\begin{equation} \tag{E2}
\begin{split}
V_{\mbox{\scriptsize Morse}}(x) \approx & A+D \lambda^2 (x-x_0)^2-D\lambda^3(x-x_0)^3  \\
 & + \frac{7}{12} D \lambda^4 (x-x_0)^4 + \cdot \cdot \cdot \quad \; .  \label{morse_potential_expansion}
\end{split}
\end{equation}
A comparison of the coefficients between Eqs.\;\eqref{simple_anharmonic_potential} and \eqref{morse_potential_expansion} leads to the following relations: $c=D\lambda^2$, $g=D\lambda^3$, and $f=\frac{7}{12}D\lambda^4$. From these relations and Eqs.\;\eqref{sound_velocity} to \eqref{debye_frequency}, the Debye temperature and the Gr\"{u}neisen constant (Eqs.\;\eqref{debye_temperature} and \eqref{Gruneisen_constant}) are written as 
\begin{equation}
\Theta_{D,0}=\frac{\hbar}{k_B} C \sqrt{c}    \tag{E3}\label{debye_temperature_2}
\end{equation}
and
\begin{equation}
\gamma_0=\frac{6}{7}x_0\frac{f}{g} \; ,  \tag{E4}\label{grunisen_constant_2}
\end{equation}
where $C$ is a constant, which does not depend on the Morse fitting parameters. Since it is assumed that $x_0$ does not contribute to thermal expansion, a normalized Gr\"{u}neisen constant is defined as
\begin{equation}
\gamma_0'=\frac{\gamma_0}{x_0}=\frac{6f}{7g} \; .  \tag{E5}\label{normalized_grunisen_constant}
\end{equation}
From the above equations, it is inferred that the lattice expansion becomes smaller or larger when the Debye temperature, $\Theta_{D,0}$, is larger or smaller and the normalized Gr\"{u}neisen constant, $\gamma_0'$, is smaller or larger.

\section*{APPENDIX F: Real-time scaling}
A correlation of relaxation time to real time is determined using the phonon component of the thermal conductivity. In this appendix, it is shown how the correlation is estimated when a system experiences an irreversible process between two equilibrium states.

The flux of thermal energy, $J$, is written from Fourier's law in one dimension as\cite{kittel1966introduction}
\begin{equation}
J = -K \frac{dT}{dx} \; ,  \tag{F1}\label{flux_energy}
\end{equation}
where $K$ is the thermal conductivity coefficient and $dT/dx$ is a temperature gradient. For a cubic sample with side lengths $L$, the flux of thermal energy and the temperature gradient are given, respectively, by
\begin{equation}
J = \frac{N}{6L^2} 3 \left(\frac{dE}{dt}\right)  \tag{F2}\label{flux_energy_2}
\end{equation}
and
\begin{equation}
\frac{dT}{dx} = \frac{T^C-T^R}{0.5L} \; ,  \tag{F3}\label{temperature_gradient}
\end{equation}
where $N$ is the number of atoms in the sample, $dE/dt$ is the total energy change rate per atom, and $T^C$ and $T^R$ are the temperatures of the center of the sample and the heat reservoir, respectively. Note that the factor 3 in Eq.\;\eqref{flux_energy_2} comes from the fact that there are three polarization modes for vibrational waves \cite{kittel1966introduction}. Furthermore, if only the phonon contribution of the thermal conductivity is considered and the rate of total energy change relative to the dimensionless time $t^*=t/\tau$ determined by the SEAQT model is used, i.e., $dE/dt^*$, then the relaxation time, $\tau$, can be derived as
\begin{equation}
\tau = \frac{N}{4LK_{\mbox{\scriptsize phonon}}} \frac{1}{T^R-T^C}\frac{dE}{dt^*} \; .  \tag{F4}\label{relaxation_time_phonon}  
\end{equation}
\\
Note that since the transient case is considered here, $T^C$ is a function of time in Eq.\;\eqref{temperature_gradient}, and $K_{\mbox{\scriptsize phonon}}$ is a function of $T^C$.

\bibliographystyle{apsrev4-1}
\bibliography{ref2}

\end{document}